# Recovering compressed images for automatic crack segmentation using generative models

Yong Huang[a,b], Haoyu Zhang[a,b,] Hui Li*[a,b], Stephen Wu*[c,d]

[a] Key Laboratory of Intelligent Disaster Prevention and Mitigation for Civil Infrastructures, Ministry of Industry and Information Technology, Harbin, Heilongjiang 150090, China

[b] Ministry of Education Key Laboratory of Structural Dynamic Behavior and Control, School of Civil Engineering, Harbin Institute of Technology, Harbin, China

[c]The Institute of Statistical Mathematics, Research Organization of Information and Systems, 10-3 Midori-cho, Tachikawa, Tokyo, 190-8562 Japan

[d]The Graduate University for Advanced Studies, SOKENDAI, 10-3 Midori-cho, Tachikawa, Tokyo, 190-8562 Japan

**Abstract**

In a structural health monitoring (SHM) system that uses digital cameras to monitor cracks of structural surfaces, techniques for reliable and effective data compression are essential to ensure a stable and energy-efficient crack images transmission in wireless devices, e.g., drones and robots with high definition cameras installed. Compressive sensing (CS) is a signal processing technique that allows accurate recovery of a signal from a sampling rate much smaller than the limitation of the Nyquist sampling theorem. The conventional CS method is based on the principle that, through a regularized optimization, the sparsity property of the original signals in some domain can be exploited to get the exact reconstruction with a high probability. However, the strong assumption of the signals being highly sparse in an invertible space is relatively hard for real crack images. In this paper, we present a new approach of CS that replaces the sparsity regularization with a generative model that is able to effectively capture a low dimension representation of targeted images. We develop a recovery framework for automatic crack segmentation of compressed crack images based on this new CS method and demonstrate the remarkable performance of the method taking advantage of the strong capability of generative models to capture the necessary features required in the crack segmentation task even the backgrounds of the generated images are not well reconstructed. The superior performance of our recovery framework is illustrated by comparing with three existing CS algorithms. Furthermore, we show that our framework is extensible to other common problems in automatic crack segmentation, such as defect recovery from motion blurring and occlusion.

*Corresponding authors.
*E-mail addresses:* lihui@hit.edu.cn (Hui Li); stewu@ism.ac.jp (Stephen Wu)

## 1. Introduction

In structural engineering, cracks are indications of local defects that may affect the vibrational response of a structure under external loadings. Traditionally, detection of cracks was manually performed through visual inspections by qualified experts. Such approach is costly and time-consuming to meet the rapidly growing need of structural health monitoring (SHM) around the world. In recent decades, development of vision sensors has allowed continuous collection of image data of civil structures (e.g., automated robots, unmanned aerial vehicles, cars armed with cameras and fixed surveillance cameras in bridges [1-6]). Coupling with modern image processing techniques and machine learning models, various structural information, such as visible damages, vibration displacements, and loads [7-9], can be automatically extracted and used to evaluate states of structural health. In particular, taking advantages of the powerful image recognition ability of deep neural networks, advanced deep learning models were used to solve different problems of automatic crack detection and segmentation from image data. For example, Xu et al. [10] proposed a steel fatigue crack detection framework for long-span bridges based on restricted Boltzmann machine (RBM), which utilizes the image classification ability of RBM to effectively identify cracks in complex backgrounds. Xu et al. [11] also proposed a novel crack detection method based on a deep fusion convolutional neural network (CNN), which can identify minor cracks at multiple scales and under complex backgrounds with high accuracy during in-field testing. Oullette et al. [12] integrated genetic algorithm with CNN to accomplish autonomous crack detection. Li et al. [7] and Ni et al. [13] studied the crack edge delineation task based on semantic segmentation. Cha et al. [14] presented a deep learning approach for multiple damages detection based on object region detection.

Despite the large number of reported successes of deep-learning-based crack detection and segmentation, most of them relied on datasets that contain high quality crack images. In practice, there are many factors that may affect the image quality. For example, data compression is often performed before transmitting the image data in order to minimize the power consumption of each data transmission because of practical constraints coming from consideration of SHM cost and limited power supply of wireless platforms, including drones and robots [15]. The more we can compress the data, the better power saving we achieve. However, perfect decompression is theoretically guaranteed only when the compression rate does not exceed the Nyquist rate. By exploiting the sparsity assumption of the signals, compressive sensing (CS) methodology is developed to overcome the signal compression limit of Nyquist rate [16-18]. Many SHM applications of the method were reported for time series data [19-21], but only a few have considered applications of SHM image data [15].

The high compression rate achieved by CS depends on the assumption of signal being sparse in an invertible space. Although it could be a realistic assumption for certain engineering problems, it is often very hard to access the level of high sparsity for a problem, or sometimes to even prove the validity of the sparsity assumption. To release from this constraint, a new approach of CS using generative models has been developed [22, 23]. The key idea is to replace the L1 regularization that represents the sparsity constraint with a generative model that is able to effectively capture a low dimension representation of the signal. As a result, the ill-posedness of the inverse problem of CS is alleviated by the dimension reduction effect coming from the generative model. Such approach to solve inverse problems has been adopted as a general framework for a diverse set of applications [24-27]. In this paper, we developed a new approach for crack segmentation using compressed image data based on the CS framework using generative models. Our method is able to achieve a higher compression rate with a higher robustness toward signal noise than well-established traditional CS methods in terms of the segmentation accuracy. We illustrated that the outstanding performance of our approach comes from the effectiveness of the generative model to capture the necessary features required in the crack segmentation task even the backgrounds of the generated images are not well reconstructed. This implies that the typically strict requirement of a large training data set to build a reliable generative model can be relaxed in our framework. Furthermore, we showed that our model is extensible to other common problems of automatic crack segmentation using image data, such as defect recovery from motion blurring and occlusion.

The rest of the paper is organized as follows. Section 2 introduces the background on CS, generative models, and the integration of the two ideas for crack segmentation. An illustrative example is presented in Section 3 with comparison to other CS algorithms using sparsity constraint. Section 4 shows some examples of the extended applications of our method, followed by some concluding remarks at the end.

## 2. Methodology

### 2.1 Compressive sensing

CS is an efficient signal processing technique to compress a signal with a sampling rate much smaller than that required by the Nyquist sampling theorem, and then to reconstruct the original signal back accurately.

In CS, the data vector **y** from the compressive sensor is composed of $M$ linear projections of the signal **s**:

$$\mathbf{y} = \mathbf{A}\mathbf{s} + \boldsymbol{\varepsilon} \qquad (1)$$

where $\mathbf{s} \in \mathbb{R}^N$ represents a complete image, which is rearranged as a vector; $\mathbf{A} \in \mathbb{R}^{M \times N}$ denotes an observation matrix that is used to compress the image data **s**, where $M \ll N$; $\boldsymbol{\varepsilon}$ represents any measurement noise, which will be relatively small.

After data compression, the dimension of **y** is significantly reduced compared with the original signal **s**. For accurate data decompression, it involves solving highly underdetermined system of equations, i.e., reconstructing the original signal **s** from **y**. To solve this ill-posed problem, the sparsity of the representation of **s** in some basis can be

exploited, i.e., $\mathbf{s}$ is represented in terms of a set of basis vectors as:

$$\mathbf{s} = \mathbf{\Psi}\mathbf{w} \qquad (2)$$

where $\mathbf{\Psi} = [\mathbf{\Psi}_1, \cdots, \mathbf{\Psi}_N]$ is the $N \times N$ basis matrix with the basis vectors $\{\mathbf{\Psi}_n\}_{n=1}^N$ as columns and $\mathbf{w}$ is the sparse coefficients or weight vector. Then the ill-posed problem can be posed as a least-squares regularization equation to estimate $\mathbf{w}$ as:

$$\widetilde{\mathbf{w}} = \arg\min_{\mathbf{w}}\{\|\mathbf{y} - \mathbf{A}\mathbf{\Psi}\mathbf{w}\|_2^2 + \lambda\|\mathbf{w}\|_1\} \qquad (3)$$

where the penalty parameter $\lambda$ scales the regularization term to penalize non-zero weight values. A series of methods have been proposed to perform CS signal reconstruction, including Basis Pursuit [28, 29] and Orthogonal Matching Pursuit [30]. The choice of an L1-norm for the regularization in Eq. (3) is important to induces sparsity in $\widetilde{\mathbf{w}}$ and perform the inversion with a high accuracy.

Note that the assumption of sparsity is the core for traditional CS reconstruction methods. The process of iterative calculation of constraint optimization and finding appropriate sparse bases in Eq. (3) may be computational demanding. More importantly, it is not guaranteed to always be able to find a very sparse representation of the original signal, especially for real image data of civil structures. Therefore, the applications of the traditional CS methods can be very limited. In this research, a new CS method using generative models is presented for two-dimensional crack image reconstruction, in which the sparsity constraint is not required.

## 2.2 Generative models

In machine learning, generative models are models that can generate new data instances or samples from a conditional probability distribution. Often, a generative model learns a lower dimension representation of the sample space and it includes many categories, such as variational autoencoder (VAE) [31, 32] or generative adversarial network (GAN) [33]. Before introducing our method, we briefly explain the generative model that will be used in this research, which is GAN. In theory, any kind of generative models can be adopted, but we have chosen GAN because of its outstanding ability to learn the low dimension representation of many different classes of images.

The key idea of GAN is to train two neural networks, a generator $G$ that produces images and a discriminator $D$ that classifies a given image to be real or fake (artificially generated), simultaneously by letting them to compete with each other in a game: $G$ tries to trick $D$ with what it generates, while $D$ tries to identify the fake images. The idea is implemented by training both networks as a minimax game using a single loss function:

$$\min_G \max_D V(D, G) = \mathbb{E}_{x \sim p_{data}(x)}[\log D(x)] + \mathbb{E}_{z \sim p_z(z)}[\log(1 - D(G(z)))] \qquad (4)$$

The solution of this optimization problem can be numerically estimated by an iterative scheme based on two loss functions $L_G$ and $L_D$ for generator $G$ and discriminator $D$, respectively, in Eq. (5). For image generation, $G$ maps an n-dimensional vector $z$ into three high dimensional matrices, where values of $z$ are typically drawn from standard Gaussian distributions. Fig. 1 shows an outline of GAN for crack images.

$$L_G = \mathbb{E}_{z \sim p_z(z)}\left[\log\left(1 - D(G(z))\right)\right] \qquad (5a)$$

$$L_D = \mathbb{E}_{x \sim p_{data}(x)}[\log(1 - D(x))] + \mathbb{E}_{z \sim p_z(z)}[\log D(G(z))] \qquad (5b)$$

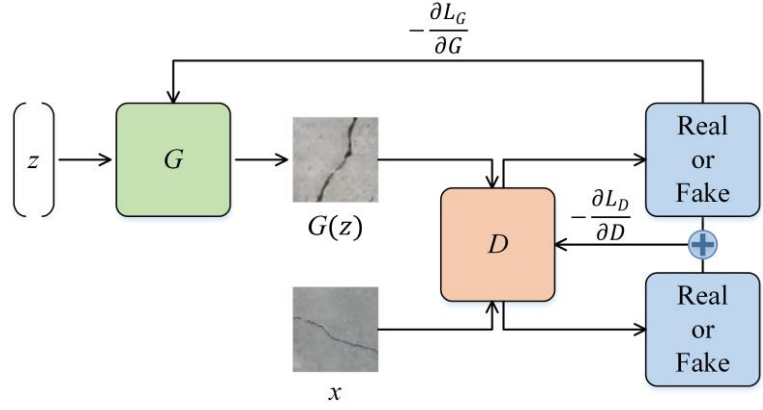

Fig. 1. The outline of training GAN for crack images

**2.3 Crack segmentation of compressed images**

Our method contains two steps, where each involves a deep neural network. In the first step, compressed image data is recovered using CS with generative model. In the conventional CS, the assumption of sparsity is used to constraint the search space of the inverse problem as shown in Eq. (3). Assuming we have successfully trained a reliable generator network $G$ for crack images, we reformulate the optimization problem based on $G$ as follow:

$$\hat{z} = \arg\min_{z}\{L_c + \lambda L_p\} \tag{6}$$

where $\lambda$ is the penalty parameter. The basic idea behind this formulation is that if $G$ is able to captures the features of the crack images using an n-dimensional vector $z$, we can directly search for optimal $\tilde{s} = G(\hat{z})$ in the $z$ space instead of the original high dimensional image space $s$. In other words, the generator $G$ is used as an implicit regularizer to replace the sparsity-based regularization. Bora et al. [22] provided theoretical justification that as long as the optimization algorithm finds a good approximate solution $\hat{z}$ in Eq. (6), the output $G(\hat{z}) = \tilde{s}$ will be the closest possible solution to the original image $s$ within the range of $G$.

Note that there is an extra penalty term $L_p$ to increase the robustness of the solution $\hat{z}$. In this study, we used a L2-norm for both $L_c$ and $L_p$, as shown in Eq. (7). Fig. 2 shows the flow of recovering a compressed image.

$$Recovery\ Loss = L_c(G(z), \mathbf{A}, \mathbf{y}) + \lambda L_p(z) \tag{7a}$$

where:

$$L_c(G(z), \mathbf{A}, \mathbf{y}) = \|\mathbf{A}G(z) - \mathbf{y}\|_2^2 \tag{7b}$$

$$L_p(z) = \|z - 0\|_2^2 \tag{7c}$$

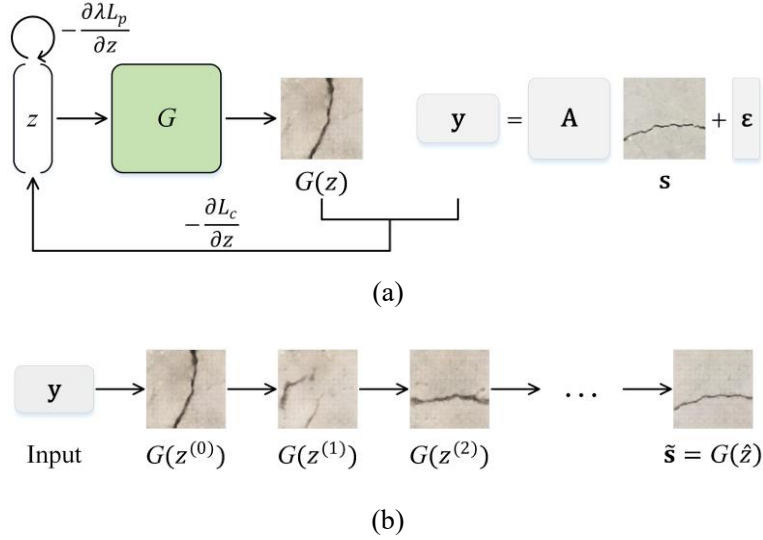

(a)

(b)

Fig. 2. The first step of proposed framework for CS data decompression using generative models. (a) Given a vector $z$ that is randomly drawn from the standard Gaussian distribution, and a generator $G$ trained for generating crack images, we iteratively update $z$ to find the optimal solution $\hat{z}$ by the recovery loss in Eq. (7). (b) The transferring of generated images $G(z^{(k)})$ when iteratively updating $z$, where $k$ denotes the $k^{th}$ iteration.

In the second step, a well pre-trained crack segmentation neural network is applied to the decompressed image. The goal of this neural network is to label each pixel in the image to be a crack or not. In this paper, we demonstrated that a typical GAN generator $G$ is already capable of capturing the important features in an image that are needed for the crack segmentation task even when the generative model is not perfectly trained. In other words, we can be relaxed from the typically strict demand of large training data set when building the generative model. Therefore, the CS method based on generative models is suitable for the automatic crack segmentation task. As will be seen in the next section, the *G* trained in this study did not do a perfect job on the recovery of the image background (i.e., non-crack area in the crack segmentation case), yet the segmentation task is well performed. In order to avoid the influence from the noise in the background, we used an extended fully convolutional network [34] that is well studied, and has the ability to accurately predict pixel-wise crack segmentation in various and complex backgrounds.

## 3. Illustrative example

We demonstrated the details of model training and algorithm implementation of the proposed method through an illustrative example using real data of cracks images. The results of our method were compared with other well-established methods and exhibited an outstanding improvement, especially in the case of noisy data.

### 3.1 Dataset and configuration of training platform

#### 3.1.1 Image data

In this study, we used Özgenel's dataset [35], which contains 20,000 crack images in different backgrounds, including various crack location and surface finishes. Note that Özgenel and Sorguç performed some pre-processing steps on the original crack images to produce the dataset. The crack images are split from 458 full resolution images (4032 pixels

× 3024 pixels) to reduce the computational demand when training and using the neural networks. The resolution of each split images are 227 pixels × 227 pixels in size, so some cracks in the dataset may look relatively wide. To further optimize the performance of model training on a standard personal computer, we used a bilinear interpolation method to resize the resolution of all crack images to 128 pixels × 128 pixels such that the resolution of input data is in the power of 2. Some of the resized images are presented in Fig. 3(a). These resized images contain a variety of cracks from wide to thin. For validation of our method, we selected 50 images consisted of a combination of various types of cracks, such as wide cracks, thin cracks, branched cracks and so on, that are excluded from the GAN model training. These images were used to compare the segmentation results among different well-established CS algorithms. Some of the validation images are shown in Fig. 3(b).

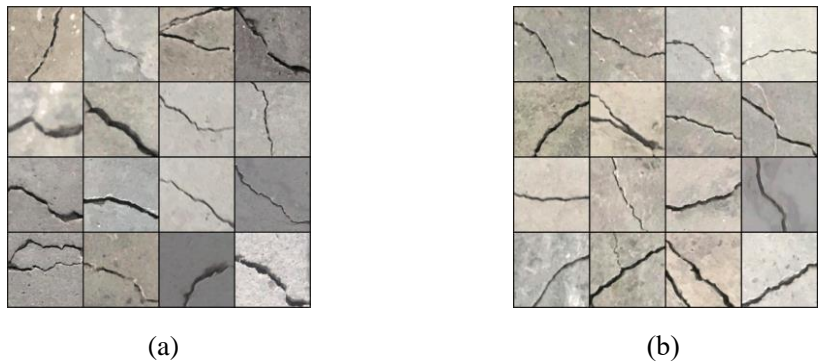

(a)                                                       (b)

Fig. 3. Example of (a) resized images and (b) validation images.

### 3.1.2 Configuration of training platform

In subsequent experiments, we use PyTorch [36] as our developing environment for implementation of training neural network models and other auxiliary computations. A personal computer with regular configurations is used for executing all computing, and its configurations are listed as follow: the operate system is Ubuntu 18.04, the central processing unit (CPU) is i7-8700K, the graphics processing unit (GPU) is NVIDIA GeForce RTX 2080, and the capacity of random access memory (RAM) is 16 gigabytes.

**3.2 Training of generative model**

In this example, we trained a generative model called Deep Convolutional GAN (DCGAN) proposed by Radford et al. [37], which uses convolutional layers as its name suggested and is one of the popular generative network designs available in the literature. We specified the size of training images and its generated images as 128 pixels × 128 pixels × 3, where 128 pixels × 128 pixels and 3 refer to the resolution of images and the number of color channels (red, green and blue), respectively. The discriminator $D$ is composed of convolutional layers, batch normalization layers, Leaky Rectified Linear Unit (ReLU) layers as the activation function of the intermediate outputs and Sigmoid layer as the activation function of the final output. Given an input of a single image of the size 128 pixels × 128 pixels × 3, $D$ will output a probability value to denote whether the input image is a "real image", i.e., not artificially generated. On the other hand, the generator $G$ consists of transposed convolutional layers, batch normalization layers, ReLU layers as the activation functions of the intermediate outputs and a Tanh layer as the activation functions of the final output. We set the input layer to have 100 neurons in our models, i.e., the input $z$ is only 100-dimentional, for which the values of each neuron are sampled from a standard normal distribution $\mathcal{N}(0,1)$. The output of $G$ is a colorful image with the size of 128 pixels × 128 pixels × 3. Note that the pixel values of the generated and real images are normalized within the interval $[-1, 1]$ when training the DCGAN. We adopted an empirical setup in the iterative training steps for improving the quality of generated images that parameters in $G$ were updated twice after parameters in $D$ were updated once. The detailed parameters and shapes of each layer of our DCGAN model are demonstrated in Fig. 4.

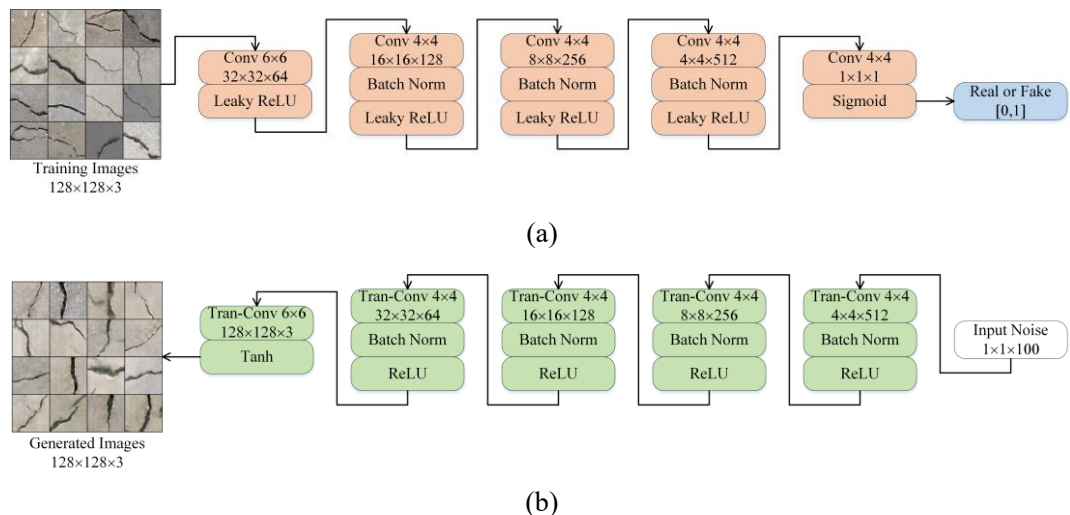

(a)

(b)

Fig. 4. Detailed architecture of (a) the discriminator $D$ and (b) the generator $G$ in DCGAN trained for generating crack images.

We selected the hyperparameters in DCGAN based on trial-and-error, and their values are listed as follow: the size of each minibatch is 16, the number of maximum training epoch is 25, the optimizers for training both the discriminator and generator are adaptive moment estimation (Adam) with a learning rate of 0.0002, and the value of betas in Adam optimizer is (0.5, 0.999). The quality of the generated crack image increased as the iterative training proceeded. Fig. 5 shows some examples of the generated images after training the DCGAN for 25 epochs. The backgrounds of the crack images are very rough comparing to the real images, but this issue does not affect the performance of crack segmentation.

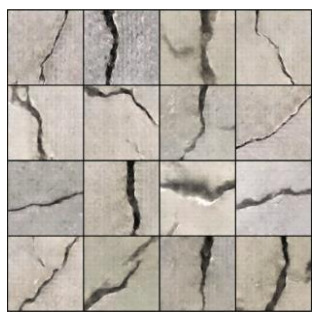

Fig. 5. Example of generated crack images after training the DCGAN for 25 epochs.

### 3.3 Parameter setup for CS using generative models

As described in Fig. 2, in our recovery framework, the input vector $z$ is randomly initialized, and the value of $z$ is then updated by the gradient descent method using Eq. (6) and Eq. (7) by fixing the network parameters of the generator $G$. In the searching of the optimal value of $\hat{z}$, the gradient descent method is implemented by Adam optimizer, the hyperparameters are selected as follows: the learning rate of Adam optimizer is 0.1, the value of betas in Adam optimizer is (0.9, 0.999), and the proportion $\lambda$ of penalty item in Eq. (7) is 0.001. The number of iterations is taken as 200 for a reconstruction process of one image, which can ensure that the loss curve of the optimization process converges. Because we observed that the chosen optimization method was sometimes trapped in local optimums, we re-run the optimization 10 times using randomly picked initial values for each task. The sensitivity of results from the number of repeated optimizations were discussed in Section 3.5.2.

### 3.4 Training of neural network for crack segmentation

At the second step of the proposed method, we need a semantic segmentation model for delineating region of cracks in images. Given a crack image to the model, its segmented results label each pixel as backgrounds or crack regions. In our study, we want to compare the effectiveness of different CS decompression methods for the crack segmentation task. Therefore, the same segmentation model is applied to the decompressed images from all decompression methods for fairness of comparison, and the model should have a high accuracy of the segmentation task on the 50 selected images that are excluded from our GAN model training.

However, training an accurate crack segmentation model from scratch requires lots of pairs of crack images and their ground truth of crack regions, which are manually segmented, and the training time is also very long. To alleviate such requirements, the following transfer learning [38] procedure was performed: we first found an extended fully convolutional network model [34] that was well trained for the crack segmentation task in a dataset with diverse crack images. Then, we manually segmented the crack regions in the 50 selected images. Finally, we fine-tuned the pre-trained model parameters with the manually labeled data using the exact same training parameters provided by Liu et al. [34].

For the fine-tuned crack segmentation model, after given a colorful image with the size of 128 pixels × 128 pixels × 3, then it will output a binary image with the same resolution, where 1 represents the "crack" class and 0 represents the "non-crack" class. Some examples of the image segmentation results are shown in Fig. 6.

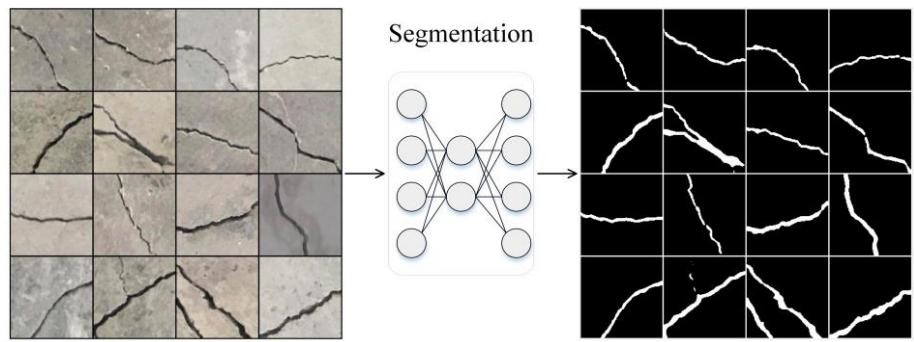

Fig. 6. Example of crack segmentation results by the fine-tuned crack segmentation model.

To evaluate performance of the fine-tuned crack segmentation model, the metrics of Accuracy and F1 score based on the indices of True Positives (TP), True Negatives (TN), False Positives (FP) and False Negatives (FN) for binary classification problems are introduced, which are defined as:

$$\text{Accuracy} = \frac{TP+TN}{TP+TN+FP+FN} \tag{8a}$$

$$F_1 = 2 \cdot \frac{\text{Precision} \cdot \text{Recall}}{\text{Precision}+\text{Recall}} \tag{8b}$$

where:

$$\text{Precision} = \frac{TP}{TP+FP} \tag{8c}$$

$$\text{Recall} = \frac{TP}{TP+FN} \tag{8d}$$

Compared with the pre-trained model, after being fine-tuned, the crack segmentation model achieves much higher average Accuracy (from 0.979 to 0.991) and average F1 score (from 0.865 to 0.937) in our dataset. The segmentation results based on the uncompressed images were manually checked and were used as the baseline for evaluating the performance of our recovery framework.

**3.5 Results**

**3.5.1 Data compression**

In our experiment, we compressed the crack images based on Eq. (1). To investigate the segmentation performance based on the compressed image, six compression ratios (CR) are selected, that is: CR = 2, 4, 8, 16, 32, and 64. The compression ratio is defined as CR = $N/M$, where $N$ and $M$ are the number of columns and rows in the observation matrix $\mathbf{A}$, respectively. For each compressive measurement, the observation matrix $\mathbf{A}$ is sampled from a standard normal distribution $\mathcal{N}(0,1)$, and zero mean Gaussian random noises $\boldsymbol{\varepsilon}$ are added after observation. The standard deviation of $\boldsymbol{\varepsilon}$ is a random variable sampled from a uniform distribution within the interval of 0 and $\delta$, where $\delta$ is the max range of values in complete image $\mathbf{s}$ multiplied by 0.05, for example, $\delta$ is 0.1 for images that are normalized within the interval $[-1, 1]$.

3.5.2 Sensitive to initial values

We observed that large image reconstruction error in terms of segmentation results might occur even when the optimized value of the loss function has converged. Hence, we randomly selected 10 images from the 50 images for validation to investigate the important factors that affect the image reconstruction accuracy. For each image, we repeat the CS reconstruction for 100 times based on the compressive measurements of a predefined CR, and the initial value of $z$ is different for each time. In each time, we record the minimum loss $L_{min}$ after optimization, and use the fine-tuned crack segmentation model to calculate the quantitative similarity metrics, i.e., Accuracy and F1 score, between the true crack regions and those in the $\tilde{s} = G(\hat{z})$ that is corresponding to the minimum loss.

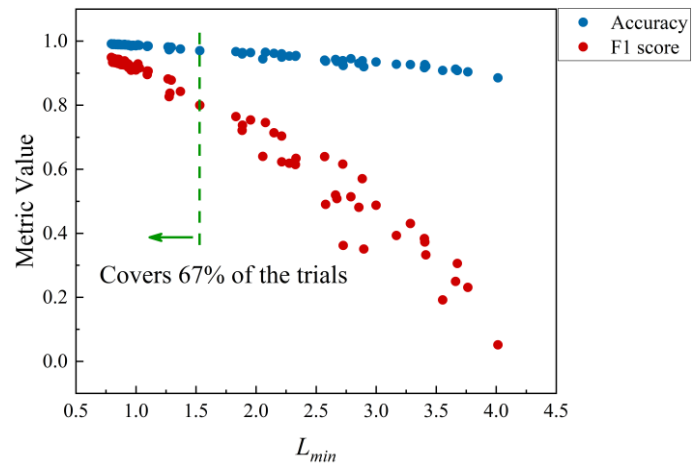

Fig. 7. The relationship between $L_{min}$ and the two quantitative similarity metrics, Accuracy and F1 score.

Note that it's an example that contains 100 trials of one image when CR = 16.

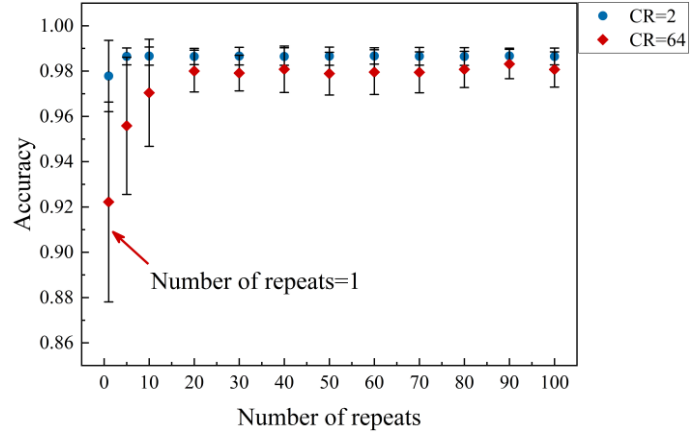

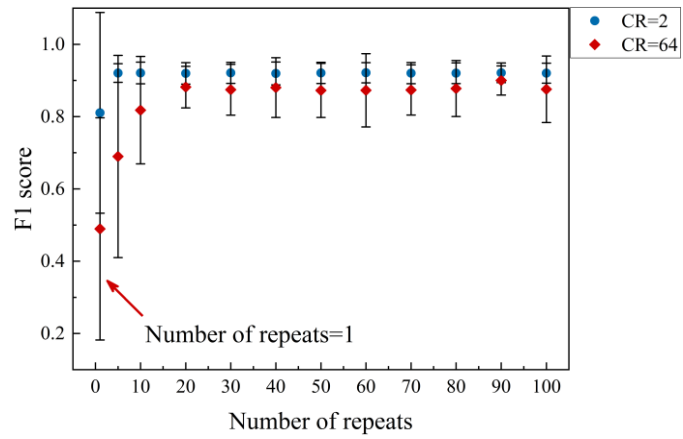

Fig. 8. The average values with error bar of (a) Accuracy and (b) F1 score, versus number of repeated optimizations, over segmentation results of the selected 10 images for two CR values.

The relationship of $L_{min}$ versus the two quantitative similarity metrics is presented in Fig. 7. The result showed that the performance of a reconstruction result is strongly correlated with the minimum loss. The smaller the loss $L_{min}$ after optimization is, the more accurate the crack segmentation results of reconstructed images will be. Therefore, we concluded that the failure of image reconstruction is mainly caused by sensitivity of output to the initial value of *z*. In other words, there are many local minimums in our problem that the optimizer failed to consistently reach the global minimum. Therefore, by using different random input *z*, we repeated the optimization of a single task for multiple times and picked the result with the lowest value of loss $L_{min}$ in order to increase the robustness of the image reconstruction. Fig. 8 shows the average Accuracy and F1 score with error bar over the selected 10 images for two CR values, which is 2 and 64. It demonstrates the effect of the number of repeated optimizations on the segmentation performance under different CR. For computational concern, we uniformly picked the number of repeated optimizations to be 10 in our study.

### 3.5.3 Comparing segmentation results

Based on the proposed experiments stated above, the image reconstruction based on the compressive measurements with six CRs are performed on the 40 crack images used in validation. For comparison, we chose three well-established traditional CS methods (BP, CoSaMP, and VB-BCS) [39-41], and performed data reconstruction experiments using the same compressive measurements. It is worthwhile to note that our method is able to reconstruct colorful images, while the three traditional CS methods is only applicable to grayscale images, which only have one color channel for brightness.

In Fig. 9, some reconstruction results of the four methods are demonstrated, when the CR is 16. It is obvious that our CS method using generative models (GAN-CS) has the highest Accuracy and F1 score.

For six CRs, the detailed distributions (including averages, standard deviations and so on) of Accuracy and F1 score between crack regions of the real and reconstructed crack images can be computed, and the results are shown as box plots in Fig. 10. It is seen that with the increase of CR, the reconstruction accuracy of the three traditional CS methods has a significant decrease, while the reconstruction accuracy of the CS method using generative models only decreases a little. This shows the superior performance of our CS method, especially when CR is high.

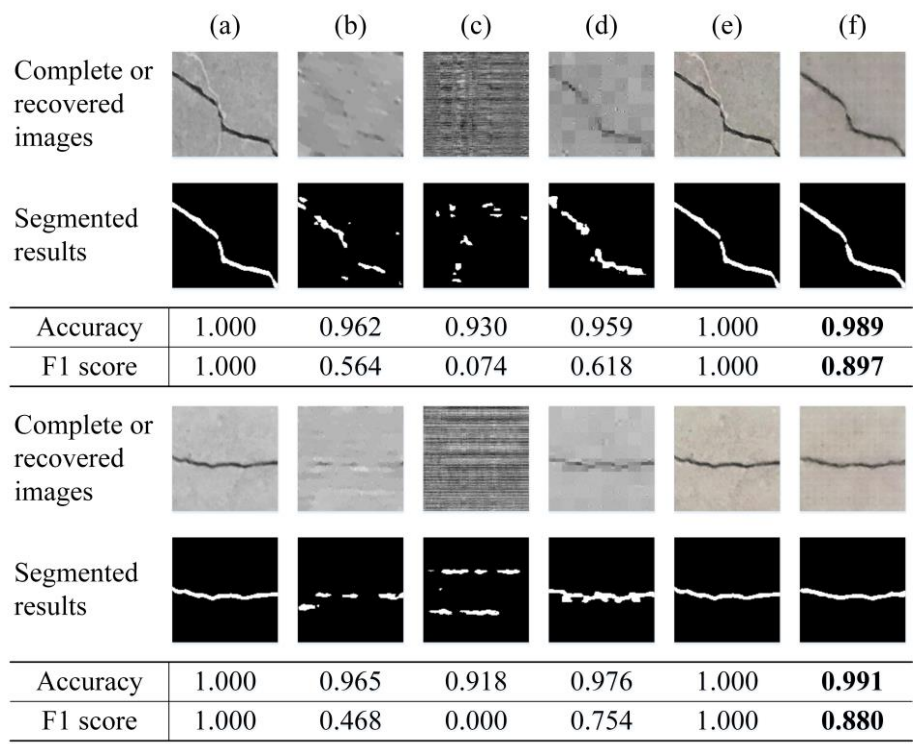

Fig. 9. The comparison of the image reconstruction performance among four CS methods, when the CR is 16. (a) complete image (grayscale); (b) image reconstructed by BP (grayscale); (c) image reconstructed by CoSaMP (grayscale); (d) image reconstructed by VB-BCS (grayscale); (e) complete image (**colorful**); (f) image reconstructed by GAN-CS (**colorful**).

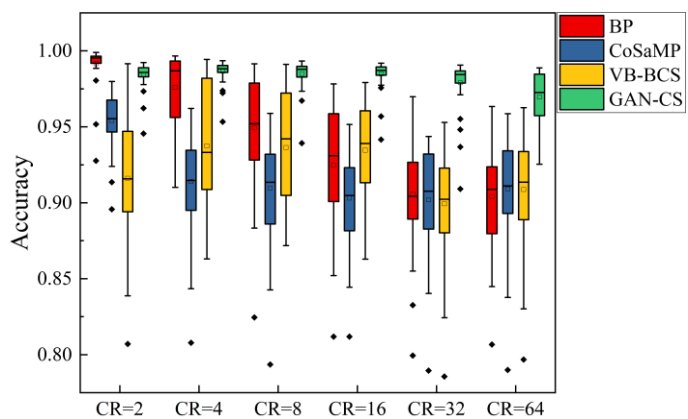

(a)

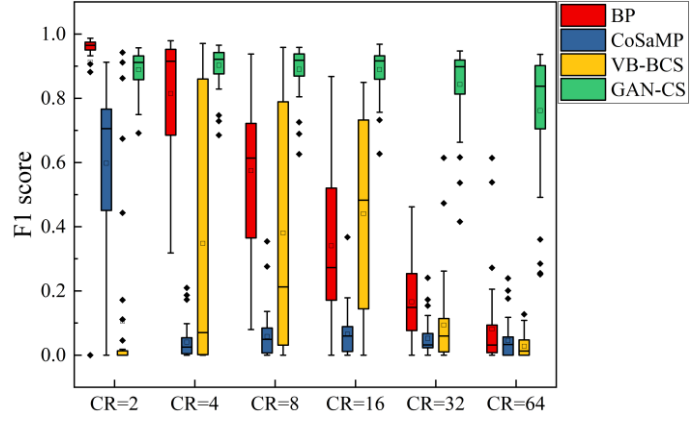

(b)

Fig. 10. The comparison of the (a) Accuracy and (b) F1 score of the four CS algorithms, at different CRs.

In Table 1, average computation times for producing one reconstructed image in six CRs are tabulated, corresponding to the results in Fig. 10. Since a GAN model is well trained before the image reconstruction, our presented method is much faster than the other three algorithms. The computation speed of our method is nearly 200 times higher than that of the slowest one (BP) even though the amount of data our method reconstruct is three times more than other methods. Overall, the comparison results in Fig. 10 and Table 1 demonstrate that our proposed CS method is superior to the three traditional CS methods in both of the calculation speed and the performance of automatic crack segmentation.

Table 1. Average computation times for producing one reconstructed image in six CRs.

| Method | BP | CoSaMP | VB-BCS | GAN-CS (Training GAN) |
|---|---|---|---|---|
| Time Cost (min.) | 203.54 | 8.38 | 59.71 | 0.94 (13.25) |

### 3.5.4 Sensitive to image noise

In Fig. 11, as the ratio between $\delta$ of random noise $\varepsilon$ and the max range of values in complete image $s$ (note that the ratio, NL, represents the level of noise $\varepsilon$) increased from 0,

0.05, to 0.1, the segmentation accuracy of our method is almost not reduced, while the segmentation accuracy of other three CS methods is significantly reduced. The results in Fig. 11 demonstrate that our recovery framework is more robust to random noise than other three traditional CS methods.

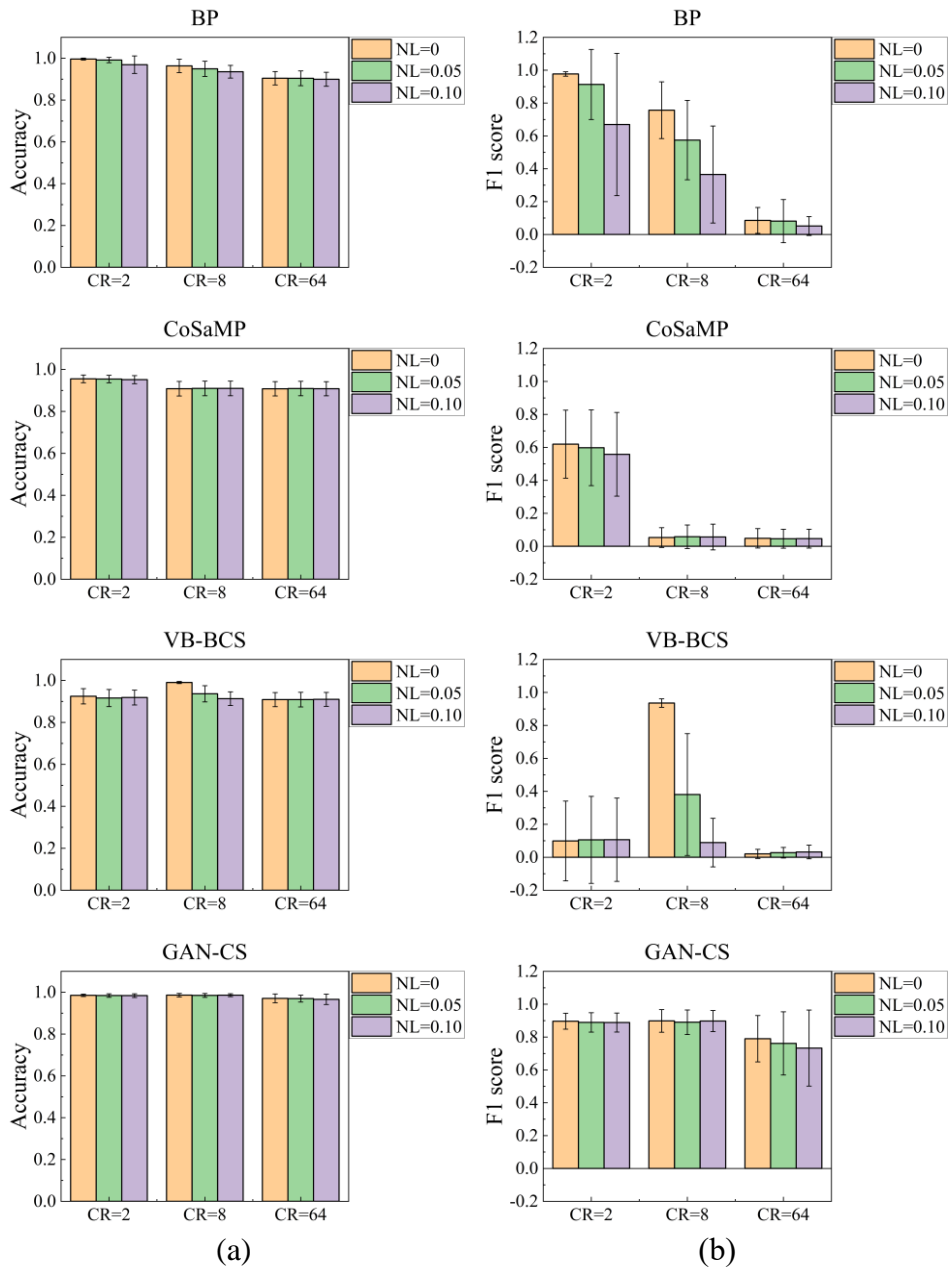

Fig. 11. The comparison of the (a) Accuracy and (b) F1 score at different CRs and noise levels (NL), over segmentation results of the four CS algorithms.

## 4. Extended applications

In addition to the application of CS data decompression, we further examine the possibility to apply our CS recovery framework to complex computer vision tasks including motion blurring removal and occlusion removal. Motion blurring of the captured images were usually caused by a mismatch between the movement speed of data acquisition devices and the shutter speed of their cameras. Meanwhile, occlusion was caused by blocking of some parts of the image due to leaves, shadows, and other debris, i.e., some valid information in the captured images was occluded. Both cases will significantly affect the performance of standard crack detection and segmentation algorithms.

Both scenarios above can be described as a data loss process with the following mathematical formulation:

$$\mathbf{y} = \mathbf{A} \odot \mathbf{s} + \boldsymbol{\varepsilon} \tag{9}$$

where $\odot$ denotes the data loss mode. In CS, it's a Hadamard product.

Conventional deep-learning-based motion deblurring removal [42, 43] and occlusion removal [44] methods requires massive paired data (e.g., motion blurring images and their clear images, images with occlusion and their complete images) to train a recovery model. However, our CS recovery framework does not require such data, while $\mathbf{A}$ in Eq. (9) becomes more complicated as compared to the simple random value matrix in the data decompression task.

### 4.1 Motion blurring removal

In computer vision, motion blurring is generated by filtering clear images with a motion blurring kernel (a small size matrix determined by two parameters, the motion direction and the degree of blurring). Therefore, now in Eq. (9), **A** is the motion blurring kernel and $\odot$ denotes the operation of filtering clear images with kernel **A**. Since the parameters of motion blurring kernels can be estimated from motion blurring images by preprocessing methods [45], we directly give motion blurring kernels and generate different motion blurring images for subsequent experiments using Kornia [46], which is a computer vision library for PyTorch [36]. In Fig. 12, the generation of motion blurring images is shown.

After applying our CS recovery framework to the motion blurring removal task for cases with different motion directions and blurring degrees, it is found that the performance of our crack segmentation model for images with motion blurring along the crack direction is almost unaffected. Thereby, we mainly investigated the case in which the direction of motion blurring is roughly orthogonal to the crack direction. In performed experiments, the ratio NL is 0, the blurring degree is 13, the number of repeat times for recovering one image is uniformly adjusted to 5, and other hyperparameters in data recovery process remain consistent with that in the application of data decompression. Some examples of motion blurring removal are shown in Fig. 13. And the average values and standard deviations of F1 score in segmentation results before and after motion blurring removal are compared in Table 2, which indicates around 10% improvement of the F1 score after motion blurring removal.

$$\text{Filter}\left(\underset{\mathbf{A}}{\blacksquare}, \underset{\mathbf{s}}{\text{[image]}}\right) = \underset{\mathbf{y}}{\text{[image]}}$$

Fig. 12. The generation of motion blurring effect.

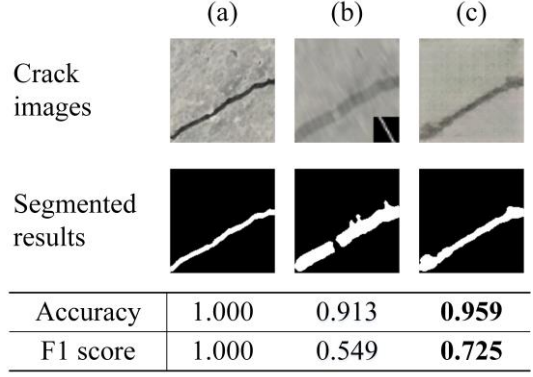

| | (a) | (b) | (c) |
|---|---|---|---|
| Accuracy | 1.000 | 0.913 | **0.959** |
| F1 score | 1.000 | 0.549 | **0.725** |

Fig. 13. The examples of motion blurring removal. (a) clear image; (b) image with motion blurring; (c) image recovered by GAN-CS.

Table 2. The comparison of F1 score before and after motion blurring removal.

| F1 score | Before removal | After removal | Growth rate (%) |
|---|---|---|---|
| Average (%) | 65.7 | 74.7 | 13.7 |
| Standard deviation (%) | 12.2 | 11.7 | -4.6 |

## 4.2 Occlusion removal

For Eq. (9) in the application of occlusion removal, **A** can be viewed as the occlusion area in images and $\odot$ denotes the operation of blending **A** with complete images. The mask region based CNN (Mask R-CNN) [47], which is a deep CNN model for automatically delineating accurate shape of multiple objects in images, is able to preprocess the selection of occlusion area. Because of this, we use a delineated occlusion area to generate partly occluded images for subsequent experiments. In Fig. 14, the procedure of occlusion area delineation and generation are presented.

We performed experiments using the 40 crack images for validation. In experiments, the ratio NL is 0, the given occlusion area is a leaf shape whose area is approximately 25% of 128 pixels × 128 pixels, the number of repeat times for recovering one image is uniformly adjusted to 5, and settings of other hyperparameters in the data recovery process remain unchanged. Some examples of occlusion removal are shown in Fig. 15. And the average values and standard deviations of F1 score in segmentation results before and after recovery are tabulated in Table 3. It is demonstrated that our recovery framework can accurately complement the occluded crack information with around 150% improvement in F1 score after occlusion removal.

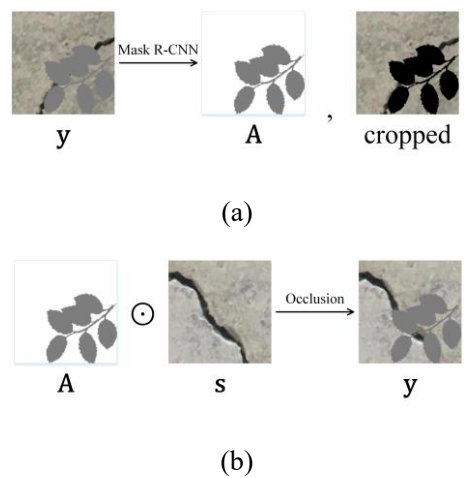

Fig. 14. The (a) delineation and (b) generation of occlusion area.

Fig. 15. The examples of occlusion removal (a) complete image; (b) image with occlusion; (c) image recovered by GAN-CS.

Table 3. The comparison of F1 score before and after occlusion removal.

| F1 score | Before removal | After removal | Growth rate (%) |
|---|---|---|---|
| Average (%) | 33.6 | 86.7 | 158.0 |
| Standard deviation (%) | 31.0 | 9.0 | -71.0 |

## 5. Conclusion

CS techniques can increase the efficiency of wireless transmission of image data in SHM systems. Most of the existing CS techniques allow effective reconstruction of images only when they are sufficiently sparse in terms of some basis. In practice, such assumption is often violated, which motivates our study to replace this traditional CS method with a new generative-model-based CS approach that is not relied on sparsity constraints and more effective for crack segmentation task in the decompressed images.

In this paper, we proposed a GAN-based CS framework for autonomously crack segmentation of compressed image data in SHM, which are composed of two steps. The first step is to train an appropriate GAN model to learn the distribution of given image dataset with cracks, so that the $G$ network of the trained GAN can generate a variety of realistic crack images. After that, the decompressed image can be obtained by solving an optimization problem in a reduced dimension space given by $G$. In the second step, a semantic segmentation model is trained to extract crack information from decompressed images, and we have shown that $G$ is able to efficiently capture useful information for the segmentation task even though the backgrounds of the generated images are not realistic enough. The experimental results show that our recovery framework can recover the original images with high accuracies in term of segmentation results, and the comparison studies show that overall the presented CS algorithm has a better performance than well-established CS algorithms. The application of the presented method is also extended to motion blurring removal and occlusion removal. Our results indicate that the proposed crack segmentation framework has the potential to be used for various image defect recovery tasks..

Relaxing sparsity constraint allows more general use of compressive sensing in various real world civil engineering applications. In addition, by using the pre-trained generative model, the computation cost for recovering an image is much smaller compared with the traditional CS methods. The method has a high potential to tackle different practical issues of automatic crack segmentation tasks in modern SHM systems.

**Acknowledgments**

This research is supported by grants from the National Natural Science Foundation of China (NSFC grant nos. 51778192, U1711265 and 51638007). Stephen Wu also gratefully acknowledges financial support from JSPS KAKENHI Grant Number JP18 K18017.